\documentclass[global,twocolumn]{svjour}
\usepackage{graphics}
\journalname{Applied Physics A}
\begin{document}
\title{Model of separated form factors for unilamellar vesicles
}
\author{M.A. Kiselev\inst{1}, P. Lesieur\inst{2}, A.M. Kisselev\inst{3}, D. Lombardo\inst{2}, V.L. Aksenov\inst{1}                 }                     

\offprints{M.A. Kiselev, kiselev@nf.jinr.ru}         
\institute{Frank Laboratory of Neutron Physics, JINR, 141980 Dubna, Moscow reg., Russia 
\and LURE, Bat. 209-D, B.P. 34, F-91898 Orsay, France 
\and Physics Department, Cornell University, Ithaca, NY 14853, USA}
\date{Received: date / Revised version: date}
%
\maketitle
\begin{abstract}
New model of separated form factors is proposed for the evaluation of small-angle neutron scattering curves from large unilamellar vesicles. The validity of the model was checked by comparison to the model of hollow sphere. The model of separated form factors and hollow sphere model give reasonable agreement in the evaluation of vesicle parameters.

\noindent{\textbf{PACS}: 87.16.Dg, 61.12.Ex, 61.10.Eq}

\end{abstract}
Information about the internal membrane structure is mainly derived from X-ray diffraction experiments on multilamellar vesicles (MLVs) \cite{Ref1}. A single lipid bilayer possesses the structure typical of most biological membranes. Unilamellar vesicles (ULVs) appear to be more biologically appealing model of the lipid membrane than multilamellar vesicles. Moreover, vesicles are used as delivery agents of drugs, genetic materials and enzymes through living cell membrane and other hydrophobic barriers \cite{Ref2,Ref3}. Today, the problem of accurate and simultaneous determination of the vesicle radius, polydispersity, and the internal membrane structure is not yet solved in SAXS and SANS experiments \cite{Ref5}-\cite{Ref9}. The information about internal membrane structure derived from the SANS experiment is based on the strip-function model of the neutron scattering length density across bilayer $\rho(x)$ \cite{Ref4} and application of the hollow sphere model for the vesicle \cite{Ref5,Ref7}. Important problem is to develop new approach to the evaluation of SANS and SAXS experimental curves with possibility to describe $\rho(x)$ as an any analytical or numerical function. The purpose of present work is to propose and verify the new analytical equations for the calculation of the SANS curves from the phospholipid vesicles.

\section{Experiment and Model}
\label{sec:1}\textbf{D}ipalmitoylphosphatidylcholine (DPPC) was purchased from Sigma (France), and $\rm{D_{2}O}$ was from Isotop (S.-Peterburg,Russia). LUVs were prepared by extrusion of MLVs through polycarbonate filter with pore diameter 500 \r{A} as described in ref. \cite{Ref7}. The spectra from unilamellar DPPC vesicles were collected at YuMO small-angle spectrometer of IBR-2 reactor (Dubna, Russia) at $T=20^{\circ}C$ \cite{Ref10}. Incoherent background was subtracted from normalized cross section of vesicles as described in ref. \cite{Ref5}. DPPC concentration in sample was 1\% (w/w).

Macroscopic cross section of monodispersed population of vesicles \cite{Ref11}
$$
\frac{d\Sigma}{d\Omega}_{mon}(q)=n\cdot A^2(q)\cdot S(q)
\eqno{(1)}
$$
where $n$ is the number of vesicles per unit volume, $A(q)$ is scattering amplitude of vesicle, and $S(q)$ is vesicle structure factor. $S(q)\approx1$ for 1\% (w/w) DPPC concentration \cite{Ref12}. Scattering amplitude $A(q)$ for the case of vesicles with spherical symmetry \cite{Ref11}
$$
A(q)=4\pi\int{\rho(r)\frac{\sin(qr)}{qr}r^2dr}
\eqno{(2)}
$$
where $\rho(r)$ is neutron contrast between bilayer and solvent. Integration o  (2) over the hollow sphere with \linebreak \mbox{$\rho(x)\equiv\Delta\rho$} leads to the hollow sphere (HS) model of the vesicle \cite{Ref11}
$$
\frac{d\Sigma}{d\Omega}_{mon}(q)=n(\Delta\rho)^2\left(\frac{4\pi}{q^3}\right)^2(A_2-A_1)^2
\eqno{(3)}
$$
where $A_i=\sin(qR_i)-(qR_i)\cos(qR_i)$, $R_1$ is inner radius of hollow sphere, $R_2= R_1+d$ is outer radius of hollow sphere, $d$ is membrane thickness.

For the bilayer with central symmetry, (2) can be rewritten as
$$
A(q)=4\pi\int_{-d/2}^{d/2}{\rho(x)\frac{\sin\left[(R+x)q\right]}{(R+x)q}(R+x)^2dx}
\eqno{(4)}
$$
Integration of (4) gives exact expression for scattering amplitude of vesicle with separated parameters $R$, $d$, $\rho(x)$
$$
\begin{array}{ccc}
A_{ves}(q)& = &4\pi\frac{R^2}{qR}\sin(qR)\int_{-d/2}^{d/2}{\rho(x)\cos(qx)dx}+\\
& + &4\pi\frac{R}{qR}\cos(qR)\int_{-d/2}^{d/2}{\rho(x)x\sin(qx)dx}\\
\end{array}
\eqno{(5)}
$$
In the case of $R\gg d/2$, $R+x\approx R$, one can obtain from (4)
$$
A_{SFF}(q)=4\pi\frac{R^2}{qR}\sin(qR)\int_{-d/2}^{d/2}{\rho(x)\cos(qx)dx}
\eqno{(6)}
$$ 
and the macroscopic cross section can be written as 
$$
\frac{d\Sigma}{d\Omega}_{mon}(q)=n\cdot F_s(q,R)\cdot F_b(q,d)
\eqno{(7)}
$$
where $F_s(q,R)$ is the form factor of a infinitely thin sphere with radius $R$ \cite{Ref9}
$$
F_s(q,R)=\left(4\pi\frac{R^2}{qR}\sin(qR)\right)^2
\eqno{(8)}
$$
and $F_b(q,d)$ is the form factor of the symmetrical lipid bilayer
$$
F_b(q,d)=\left(\int_{-d/2}^{d/2}{\rho(x)\cos(qx)dx}\right)^2
\eqno{(9)}
$$
Eqs. (7)-(9) present a new model of separated form factors (SFF) of the large unilamellar vesicles. SFF model has advantage relative to the HS model due to possibility to describe the internal membrane structure via presentation of $\rho(x)$ as an any integrable function. The approximation of neutron scattering length density across the membrane with a constant \mbox{$\rho(x)\equiv\Delta\rho$} is far from being realistic \cite{Ref4,Ref5,Ref7}, but gives possibility to make comparison of HS and SFF models. In the approximation of \mbox{$\rho(x)\equiv\Delta\rho$}, (9) is integrated to the expression 
$$
F_b(q,d)=\left(\frac{2\Delta\rho}{q}\sin\left(\frac{qd}{2}\right)\right)^2
\eqno{(10)}
$$
In present study, vesicle polydispersity was described by nonsymmetrical Schulz distribution \cite{Ref13}
$$
G(R)=\frac{R^m}{m!}\left(\frac{m+1}{\bar{R}}\right)^{m+1}\exp\left(-\frac{(m+1)R}{\bar{R}}\right)
\eqno{(11)}
$$
where $\bar{R}$ is the average vesicle radius. The polydispersity of vesicles was characterized as relative standard deviation of vesicle radius $\sigma=\sqrt{\frac{1}{m+1}}$.

Experimentally measured macroscopic cross section $d\Sigma/d\Omega$ was calculated via convolution of the $d\Sigma/d\Omega_{mon}$ with the vesicle distribution function $G(R)$ by integration over the vesicle radius from $R_{min}=110$ \r{A} to $R_{max}=540$ \r{A} 
$$
\frac{d\Sigma}{d\Omega}=\frac{\int\limits_{R_{min}}^{R_{max}}{\frac{d\Sigma}{d\Omega}_{mon}(q,R)G(R)dR}}{\int\limits_{R_{min}}^{R_{max}}{G(R)dR}}
\eqno{(12)}
$$

Finally, $d\Sigma/d\Omega$ values were corrected for the resolution function of the YuMO spectrometer as described in ref. \cite{Ref14}. 

The parameter
$$
R_f=\frac{1}{N-3}\sum_{i=1}^{N}{\left(\frac{\frac{d\Sigma}{d\Omega}(q_i)-\frac{d\Sigma}{d\Omega}_{exp}(q_i)}{\frac{d\Sigma}{d\Omega}_{exp}(q_i)}\right)^2}
\eqno{(13)}
$$
was used as a measure of fit quality, $N$ here is a number of experimental points.

\section{Results and Discussion}
\label{sec:2}
\begin{figure}

\resizebox{0.45\textwidth}{!}
{
  \includegraphics{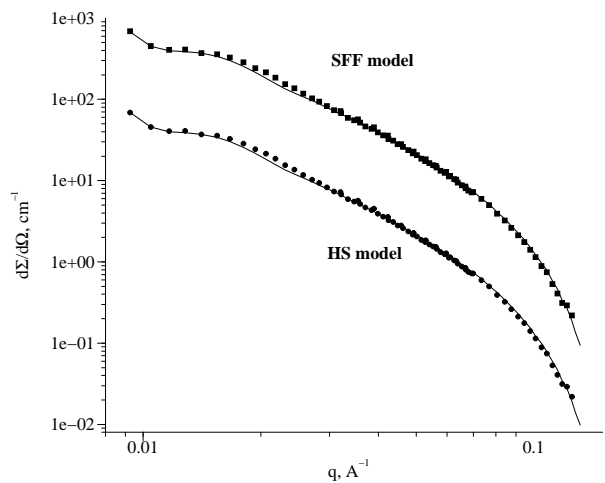}
}

\caption{SANS curves from DPPC vesicles at $T=20^{\circ}C$. Experiment (dots, squares), model calculation (solid line). Macroscopic cross sections for the calculations with HS model are in absolute units. Macroscopic cross sections for the calculations with SFF model are multiplied by 10}
\label{fig:1}  
\end{figure}

The validity of SFF model comparing to HS model was examined in the approximation of \mbox{$\rho(x)\equiv\Delta\rho$}. Fig.~\ref{fig:1} presents experimentally measured coherent macroscopic cross-section of DPPC vesicles and fitted model curves. The SFF model was applied via (7),(8),(10),(12), and the model of hollow sphere via (3),(12). As it is seen from Fig.~\ref{fig:1}, both models describe the experimental curve well enough. Free parameters used in the fit were: average vesicle radius $\bar{R}$, membrane thickness $d$, and parameter $m$ in (11).

The results of calculations are presented in Table~\ref{tab:1}. Both HS and SFF models fit experimental curve with the same accuracy, the difference in the value of the $R_f$ parameter is negligibly small, 1.3\%. HS model gives larger value of polydispersity $(\sigma =0.24)$ relative to that of SFF model $(\sigma=0.22)$, the difference is 9\%. HS model gives smaller value of average radius, the difference in radius value is 8\%. Though HS model provides the exact solution, the results of SFF model in the evaluation of vesicle radius and polydispersity do not differ more than 10\%. Important result is the same value of the calculated membrane thickness $d$ for both models. The proposed SFF model for the evaluation of SANS spectra from large unilamellar vesicles has a fundamental advantage over the model of hollow sphere. In a framework of hollow sphere model one can describe the inner structure of the membrane only in terms of a system of several inclusive concentric spheres, each having a constant scattering length density \cite{Ref5,Ref7}. The problem of water distribution function inside the lipid membrane, particularly in the region of polar head groups, is being widely discussed now. In first approximation one can use linear or exponential distribution of water from the membrane surface further inside the bilayer. This kind of water distribution will generate linear or exponential term in the function of scattering length density, which is beyond the capability of the HS model, based only on the strip-function distribution of scattering length density. The model of separated form factors introduced in the present work is deprived of this imperfection, because any integrable analytical or numerical function can be used as a function of scattering length density (see. eq. 9). Future investigation of the internal membrane structure via application of SFF model can give new interesting results for binary phospholipid/water, ternary phospholipid/cryoprotector/water or  phospholipid/surfactant/water systems.
\section{Conclusions}
\label{sec:3}
New model of separated form factors (SFF) is proposed for large unilamellar vesicles. SFF model gives an opportunity to analyze vesicle geometry and internal membrane structure separately. The validity of SFF model was examined by comparison with hollow sphere (HS) model for large unilamellar vesicles. Both models give the same value of membrane thickness, the difference in the value of vesicle average radius and vesicle polydispersity is inside of 10\% accuracy. SFF model is proposed as prospective method of the internal membrane structure evaluation from the SANS experiment on large unilamellar vesicles. 
%
%
%
\begin{table}
\caption{Vesicle parameters calculated from hollow sphere model (HS) and model of separated form factors (SFF). $\bar{R}$ is the average vesicle radius, $\sigma$ is the relative standard deviation of radius, $d$ is the membrane thickness, $R_f$ is the measure of fit quality}
\label{tab:1}       
\begin{tabular}{ccccc}
\hline\noalign{\smallskip}
Model & $\bar{R}$, \r{A} & $\sigma$ & $d$, \r{A} & $R_f$\\
\noalign{\smallskip}\hline\noalign{\smallskip}
HS & $252\pm 2$ & 0.24 & $42.6\pm 0.2$ & $0.00597$\\
SFF & $274\pm 2$ & 0.22 & $42.6\pm 0.2$ & $0.00605$\\
\noalign{\smallskip}\hline
\end{tabular}
\end{table}

\end{document}